\documentclass[aip,reprint]{revtex4-2}

\usepackage{graphicx} 
\usepackage{lastpage}
\usepackage{float}
\usepackage{fancyhdr}
\usepackage{fnpos}
\usepackage[english]{babel}
\usepackage{array}
\usepackage{droidsans}
\usepackage{charter}
\usepackage[T1]{fontenc}
\usepackage[usenames,dvipsnames]{xcolor}
\usepackage{setspace}
\usepackage[compact]{titlesec}
\usepackage{hyperref}
\usepackage{ulem}
\usepackage{braket}
\usepackage{siunitx}

\DeclareSIUnit\angstrom{\text{\AA}}

\begin{document}

\title{Self-interaction error induces spurious charge transfer artefacts in core-level simulations of x-ray photoemission and absorption spectroscopy of metal-organic interfaces}

\author{Samuel J. Hall}
\author{Benedikt P. Klein}
\author{Reinhard J. Maurer}
\email{r.maurer@warwick.ac.uk}
\affiliation{%
	Department of Chemistry, University of Warwick, Coventry, UK}%

\begin{abstract}
	First principles simulation of x-ray photoemission spectroscopy (XPS) is an important tool in the challenging interpretation and assignment of XPS data of metal-organic interfaces. We investigate the origin of the disagreement between XPS simulation and experiment for the azulene molecule adsorbed on Ag(111). We systematically eliminate possible causes for this discrepancy, including errors in the structural model and finite size effects in periodic boundary conditions. By analysis of the electronic structure in the ground-state and the core-hole excited-state, we are able to trace the error back to artificial charge transfer between adsorbed molecule and metal surface. This is caused by the self-interaction error of common exchange-correlation functionals. This error is not  remedied by standard hybrid or range-separated hybrid functionals. We employ an \textit{ad hoc} self-interaction error correction based on molecular orbital projection, that exposes this issue and is able to recover the correct experimental behaviour. The charge transfer artefact also negatively affects the prediction of X-ray absorption spectra for this system. Both the simulated photoemisson and x-ray absorption spectra show a better agreement with experimental data, once the \textit{ad hoc} correction is employed. Similar core-hole-induced charge artefacts may affect core-level simulations at metal-organic interfaces more generally.
\end{abstract}

\maketitle

\section{Introduction}

X-ray photoemission spectroscopy (XPS) plays a central role in the characterisation of chemical composition, structure, and oxidation state of materials.\cite{nilsson_applications_2002,bagus_revisiting_2019} It has proven particularly useful for the analysis and interpretation of chemical adsorption at surfaces due to its low penetration depth and high surface sensitivity. However, the changes in XP spectra associated with the surface adsorption of organic molecules or chemical changes due to surface reactions are often challenging to interpret when multiple chemical environments of the same species are present.\cite{biesinger_resolving_2011}  

First principles simulation of core level spectroscopy has proved to be an effective tool to disentangle complex overlapping XP signatures of hybrid organic-inorganic interfaces for both monolayer\cite{taucher_understanding_2016} and multilayer thin films.\cite{klein_topology_2021} Recent example studies where core-level simulation was vital for the interpretation of experimental spectra include oxygen species on Pt(111),\cite{zeng_characterization_2016} and Porphin on Ag(111) and Cu(111)\cite{diller_interpretation_2017}. Klein \textit{et al.} have shown in a series of recent joint experimental and computational studies how XPS and Near-Edge X-ray Absorption Spectroscopy (NEXAFS) provide evidence for the differences in surface chemical bonding between Naphthalene (Nt) and its topological isomer Azulene (Az) adsorbed on Ag(111), Cu(111) and Pt(111) surfaces.\cite{klein_molecular_2019,klein_molecule-metal_2019,kachel_chemisorption_2020,klein_enhanced_2020}

Despite these successes in computational simulation of XP spectra, we may ask: How do we objectively assess if a first principles simulation of a spectrum composed of so many similar signals provides the ``right prediction for the right reason'', \textit{i.e.} how do we know that it reproduced the spectrum by predicting the correct binding energies of each individual atom? This question is not easily answered for large $\pi$-conjugated molecules that feature many sp$^2$-hybridised carbon atoms with similar chemical environments. In many cases, the experimental resolution will not allow us to distinguish features with subtle chemical shifts and what is present in the experimental XP spectrum is a single broad, composite feature that incorporates most C 1s peaks. In such a scenario, chemical changes due to weak surface adsorption may lead to shifts between individual features, but the overall XPS signal may appear ultimately unaffected. Equally, errors in core-hole constrained Density Functional Theory (DFT) simulations of XP spectra may lead to the incorrect ordering of shifts between different carbons, yet the composite signal may appear to have the right shape and position. This cannot be resolved without recourse to higher-level theory such as for example Many Body Perturbation Theory at the GW level,\cite{golze_gw_2019,golze_accurate_2020} which is currently computationally intractable to be employed for molecule-metal interfaces.

For most systems, relative binding energy shifts between carbon atoms will be too small to experimentally resolve them, so this question has mostly academic value. However, naphthalene (Nt) and azulene (Az) adsorbed on Ag(111) and Cu(111) provide interesting cases where we can assess simulation accuracy. Both molecules are isomeric bicyclic hydrocarbons with the chemical formula of C$_{10}$H$_{8}$, but while naphthalene consists of two 6-membered rings, azulene possesses one 5-membered and one 7-membered ring (see inset in Fig.~\ref{fig:1}a). This seemingly small change in structure comes with a change of topology, leading to drastic differences in electronic structure, fundamental gap, reactivity, and XP spectroscopic fingerprint observed in the C 1s spectra.\cite{michl_why_1976,xia_breakdown_2014,klein_molecular_2019} In addition, the difference in topology of those two molecules leads to significantly different adsorption behaviour on metal surfaces.\cite{klein_molecular_2019,klein_molecule-metal_2019,kachel_chemisorption_2020,klein_enhanced_2020}

In the past, we have created structural models that agree with experimental structure characterisation and we have performed XP simulations of naphthalene and azulene that generally agree well with experiment. The notable exception here is the case of azulene adsorbed on Ag(111). For this system, the experimental C 1s peak shows a shoulder at low binding energies (similar to what was observed in the molecular multilayer), yet our calculations have shown no evidence of such a shoulder (see Fig.~\ref{fig:1}d). This represents a clear qualitative disagreement between theory and experiment, yet the adsorption structure is fully understood. 

In this work, we investigate the origin of the disagreement between XPS simulation and experiment for azulene on Ag(111). We systematically exclude possible causes for this discrepancy including errors in the structural model and finite size effects in periodic boundary conditions. By analysis of the electronic structure in the ground-state and the core-hole excited state, we are able to trace the error back to artificial charge transfer that arises from the self-interaction error of common exchange-correlation functionals. Crucially, this artificial charge transfer only materialises in the presence of a core-hole. This error does not appear to be fully remedied by standard hybrid or range-separated hybrid functionals. We employ an \textit{ad hoc} self-interaction error correction based on molecular orbital projection,\cite{maurer_excited-state_2013,maurer_first-principles_2014,muller_interfacial_2016} that exposes this issue and is able to describe the correct experimental behaviour. In addition, the charge transfer artefact negatively affects the prediction of NEXAFS spectra for this system and the \textit{ad hoc} correction also succeeds in improving the agreement between simulated and experimental NEXAFS data. 

\section{Computational Details}

For this work, we use periodic slab structures, the optimisation of which is described in the previous publications, where they were first described.\cite{klein_molecular_2019,klein_molecule-metal_2019} In brief, the structures consist of 4-layer ($2\sqrt{3}\times2\sqrt{3}$)-R\ang{30} metal slabs totalling 48 metal atoms in an hexagonal unit cell with a \SI{30}{\angstrom} vacuum layer. Structures were optimised with the PBE exchange correlation (xc) functional\cite{perdew_generalized_1996} and the D3 van der Waals dispersion correction,\cite{grimme_consistent_2010,becke_density-functional_2005,grimme_effect_2011} with the bottom two metal layers frozen to keep them at their bulk optimised positions, whilst the top two layers and the molecule were allowed to relax.\cite{klein_molecular_2019,klein_molecule-metal_2019}

XPS simulations were carried out using the $\Delta$SCF method\cite{gunnarsson_exchange_1976,von_barth_local-density_1979,von_barth_dynamical_1980} with binding energies obtained as the total energy difference between a ground-state and core-hole excited state calculation. Calculations were performed with periodic unit cell models and cluster model cut-outs from the repeating unit cells to leave a single molecule on the surface. 

The plane wave pseudopotential electronic structure code CASTEP\cite{clark_first_2005} was used to perform calculations with the periodic models. Calculations made use of the PBE functional,\cite{perdew_generalized_1996} a cut-off energy of \SI{450}{\eV} along with a $6\times6\times1$ k-grid ($\Gamma$-point only for free molecules and free standing periodic overlayers without a metal surface) was used with a total energy per atom convergence of at least \SI{1e-6}{\electronvolt/atom}. The core-hole was localised onto a specific atom through the creation of a modified version of default on-the-fly generated ultrasoft pseudopotential (PP),\cite{gao_theory_2008,gao_core-level_2009} with an electron removed from a core-state and the resulting charge compensated by the introduction of a positive homogeneous background charge.\cite{mizoguchi_first-principles_2009} XPS binding energies were calculated for all individual atoms in the molecule. The total XP spectrum is calculated by broadening and summation of all atom contributions. To serve as direct reference in the periodic calculations, an isolated molecule in a vacuum box to represent the `gas-phase' structure. The size of this cube was \SI{20}{\angstrom} in all directions which has previously been shown to provide a converged result.\cite{klein_nuts_2021}  

Cluster calculations were performed with the all-electron numeric atomic orbital code FHI-aims.\cite{blum_ab_2009} Calculations made use of either the \textit{force\_occupation\_projector} (FOP) or \textit{force\_occupation\_basis} (FOB) keywords to constrain a core-hole onto a specific atom. Both approaches yield numerically identical results if the core hole can successfully be localised. In some cases, localisation is easier to achieve with one or the other as previously discussed.\cite{klein_nuts_2021} FOP calculations followed the methodology set out by Kahk and Lischner\cite{kahk_accurate_2019} which involved (i) an initial calculation with an additional 0.1 electron charge on the chosen atom for 1 s.c.f step in order to break the symmetry and to localise the core-hole onto the correct atom, and (ii) removal of the extra core charge before running a full core-hole calculation. Additional core augmentation basis functions were added onto the core-excited atom to better represent the core-hole whilst a `tight-tier2' basis set was used for the molecule and first metal layer of the cluster with all lower metal layers utilising a `light-tier1' basis set. In addition to the PBE functional,\cite{perdew_generalized_1996} XPS binding energy were also calculated using the meta-GGAs SCAN \cite{sun_strongly_2015} and TPSS.\cite{tao_climbing_2003}

XPS calculation of the azulene on Ag(111) cluster cut-out with the HSE06 hybrid functional\cite{heyd_hybrid_2003} with an $\omega$ value of 0.11 bohr$^{-1}$ were carried out in FHI-aims with the FOB keyword. For the molecule a `tight-tier2' basis set was used and the metal surface was modeled with an `intermediate-tier1' basis set, the additional core augmentation basis functions as proposed by Kahk and Lischner\cite{kahk_accurate_2019} were applied to all carbon atoms involved in the structure instead of only to the excited carbon. Both FOP and FOB calculations were performed to an electronic convergence setting of 1$\times10^{-4}$e/\AA{}$^3$ for the electron density, \SI{1e-2}{\eV} for the KS-eigenvalues and \SI{1e-6}{\eV} for the total energy. Molecular Orbital projected Density-of-States (MODOS) calculations were performed in CASTEP,\cite{maurer_excited-state_2013} where the molecular orbitals (MOs) of the free standing overlayer of the system (metal adsorbed structure with the metal surface removed) are projected onto the electronic structure of the relaxed full system. 

All simulated spectra were obtained by combining the calculated core-electron binding energies with a Pseudo-Voigt broadening scheme.\cite{schmid_new_2014,schmid_new_2015,klein_nuts_2021} For the XP spectra, a full width half maximum (FWHM) value of \SI{0.7}{\electronvolt} and a Gaussian/Lorentzian (G/L) ratio of \SI{70}{\percent}/\SI{30}{\percent} was used. For the NEXAFS spectra, a photon energy dependent broadening was used based on three energy ranges with different broadening parameters. The first energy range uses a FWHM of \SI{0.75}{\electronvolt} and a G/L of \SI{80}{\percent}/\SI{20}{\percent} and ends \SI{5}{\electronvolt} above the first peak. The third range starts \SI{15}{\electronvolt} higher than the first peak and uses a FWHM value of \SI{2.0}{\electronvolt} and a G/L ratio of \SI{20}{\percent}/\SI{80}{\percent}. The intermediate range connects the two other ranges and encompasses a linearly increasing FWHM and linearly changing G/L ratio. For more detailed information about these broadening schemes see Ref.\cite{klein_nuts_2021}

\section{Results}

\subsection{Accuracy of C 1s XPS simulations of azulene and naphthalene metal-organic interfaces}

\begin{figure}[h]
    \centering
    \includegraphics{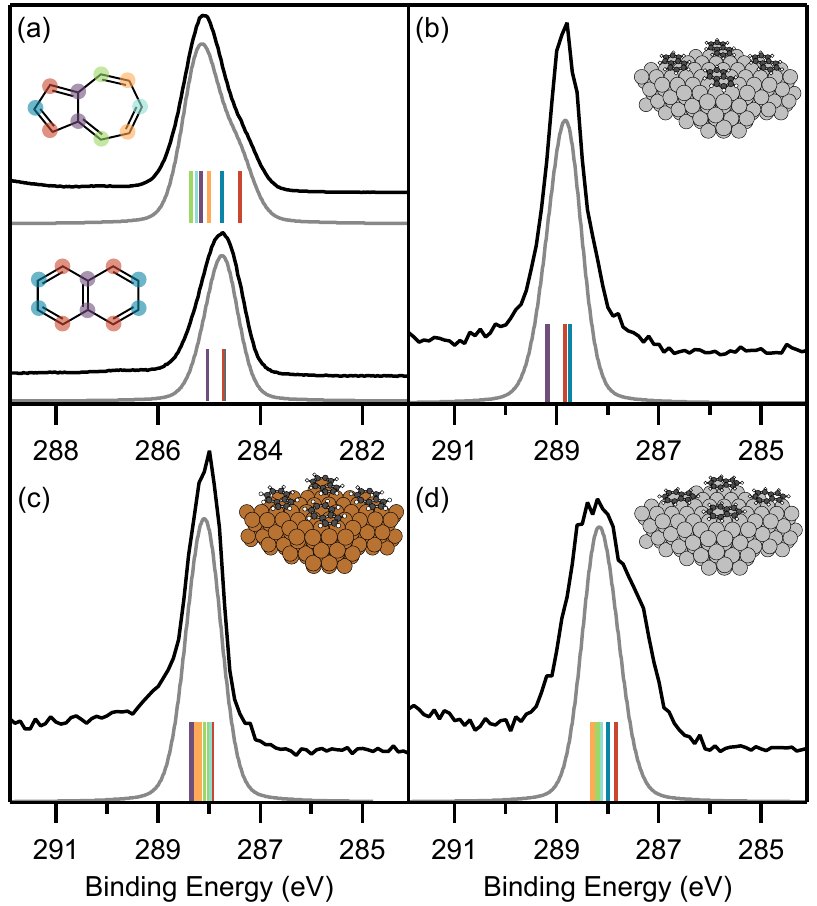}
    \caption{Comparison of experimental and simulated XPS of adsorbed azulene (Az) and naphthalene (Nt). (a) Experimental spectra of adsorbed Az and Nt multilayers\cite{klein_molecular_2019} compared against gas-phase simulations. (b) Nt adsorbed on Ag(111), (c) Az adsorbed on Cu(111), (d) Az adsorbed on Ag(111). Black lines represent experimental XP spectra. For the simulations, colored sticks represent contributions from single carbon atoms as shown in (a), with the summation included as a broadened peak in grey. 
    Experimental data taken from Ref.\cite{klein_molecule-metal_2019}}
    \label{fig:1}
\end{figure}

As part of a joint experiment-theory collaboration studying aromatic molecules adsorbed on metal surfaces, we have recently made heavy use of XPS and NEXAFS simulations to complement experimental measurements and  to provide additional insight into interaction mechanisms. In these previous publications, surface science experiments were performed for azulene and naphthalene in the multilayer\cite{klein_molecular_2019} and adsorbed on the (111) surfaces of Copper\cite{klein_molecular_2019,klein_molecule-metal_2019,kachel_chemisorption_2020}, Silver\cite{klein_molecule-metal_2019,kachel_chemisorption_2020}, and Platinum\cite{klein_enhanced_2020}. We found good agreement of our DFT calculations with the experimental values for adsorption height and energies as measured by x-ray standing wave and temperature programmed desorption as well as adsorption calorimetry experiments. In addition, these publications showed that XPS and NEXAFS simulations can achieve highly accurate predictions of experimental spectra.

In particular, DFT simulations were able to shed light onto the peculiar shape of the C 1s XPS peak of azulene. Even though the azulene molecule contains no hetero-atoms and possesses no functional groups, the C 1s peak still shows a pronounced shoulder at the low binding energy side, which is a direct result of azulene's nonalternant topology. The nonalternant topology causes an unequal charge distribution, with electron accumulation on the 5-membered ring and electron depletion on the 7-membered ring. The result is the large dipole moment (0.8 D) of azulene,\cite{tobler_microwave_1965} as well as a chemical shift in the C 1s binding energies, leading to the shoulder in the XPS peak. Our calculations showed, in agreement with earlier simulations\cite{chong_density_2010}, that the shoulder is indeed caused by the lower binding energy of the negatively charged carbon atoms in the 5-membered ring, compared to the  more positively charged atoms in the 7-membered ring. (see Fig.~\ref{fig:1}a).\cite{klein_molecular_2019} 

One could argue that the experimental spectra were recorded on a molecular multilayer of the molecule, while our calculations were performed on a theoretical model of the gas-phase molecule. However, the weak interaction between the molecules in the multilayer only exerts a minor influence on the XPS binding energies, as we could prove in a recent publication, where we performed additional calculations for the molecular crystal.\cite{klein_nuts_2021} 

Naphthalene, on the other hand, shows a quite narrow peak, because all carbon atoms have similar C 1s binding energies due to its alternant topology and homogeneous charge distribution. As the direct comparison between simulations and experimental data shows (Fig.~\ref{fig:1}a), the resolution in the measurements for the molecular multilayers is not good enough to resolve the small difference in chemical shifts between the secondary and tertiary carbon environments present in naphthalene. However, this small shift is real and can be resolved with high resolution measurements performed on molecular vapours in the gas-phase.\cite{minkov_core_2004} 

When naphthalene is adsorbed on the Ag(111) surface, the XP spectrum changes very little. Both in the measurements and the simulations, the same narrow peak remains and no differential chemical shift is introduced by the interaction with the surface (Fig.~\ref{fig:1}b). This finding is in agreement with all other experimental and computational data, which points to a weak interaction between surface and molecule and no significant charge transfer.\cite{klein_molecule-metal_2019,kachel_chemisorption_2020}

For azulene adsorbed on Cu(111), the situation is different (Fig.~\ref{fig:1}c). Here, all carbon atoms suffer a significant chemical shift, the resulting peak is residing at lower binding energy and possesses a significant asymmetry on the high binding energy side. The asymmetric peak shape can be explained by energy loss effects of the photoelectron, which are only possible in the presence of significant electronic density of states close to the Fermi energy.\cite{doniach_many-electron_1970} This energy-loss effect can not be reproduced at our level of theory. However, the calculations show significant relative shifts for all carbon atoms, and the loss of the difference in binding energies, which is caused by the inhomogeneous electron distribution of the gas-phase molecule. Therefore, XPS measurements and simulations for azulene on Cu(111) are in agreement in showing a strong interaction with the surface, which is also supported by all other experimental and theoretical data.\cite{klein_molecular_2019,klein_molecule-metal_2019,kachel_chemisorption_2020}

Thus, for both naphthalene on Ag(111) and azulene on Cu(111) it can be said that the XPS simulations compare well to the experimental XP spectra and fit perfectly into a picture of molecule-metal interaction painted by all other data. For azulene adsorbed on Ag(111), the situation is quite different. Here the experimental C 1s spectrum shows a retention of the shoulder which seems if anything more pronounced, while the DFT simulation shows a significant change in the chemical shifts and a loss of the shoulder (Fig.~\ref{fig:1}d). All other experimental and computational data points to a weak interaction between surface and molecule and no significant charge transfer,\cite{klein_molecule-metal_2019,kachel_chemisorption_2020} also leading to the expectation that the chemical shift of all carbon atoms should not significantly change due to adsorption and the peak shoulder should be retained. In the following, we will investigate the cause of the missing shoulder in the simulated C 1s XPS data for azulene on Ag(111).

\subsection{The case of the missing XPS shoulder for azulene on Ag(111)}

\begin{figure}[h]
    \centering
    \includegraphics{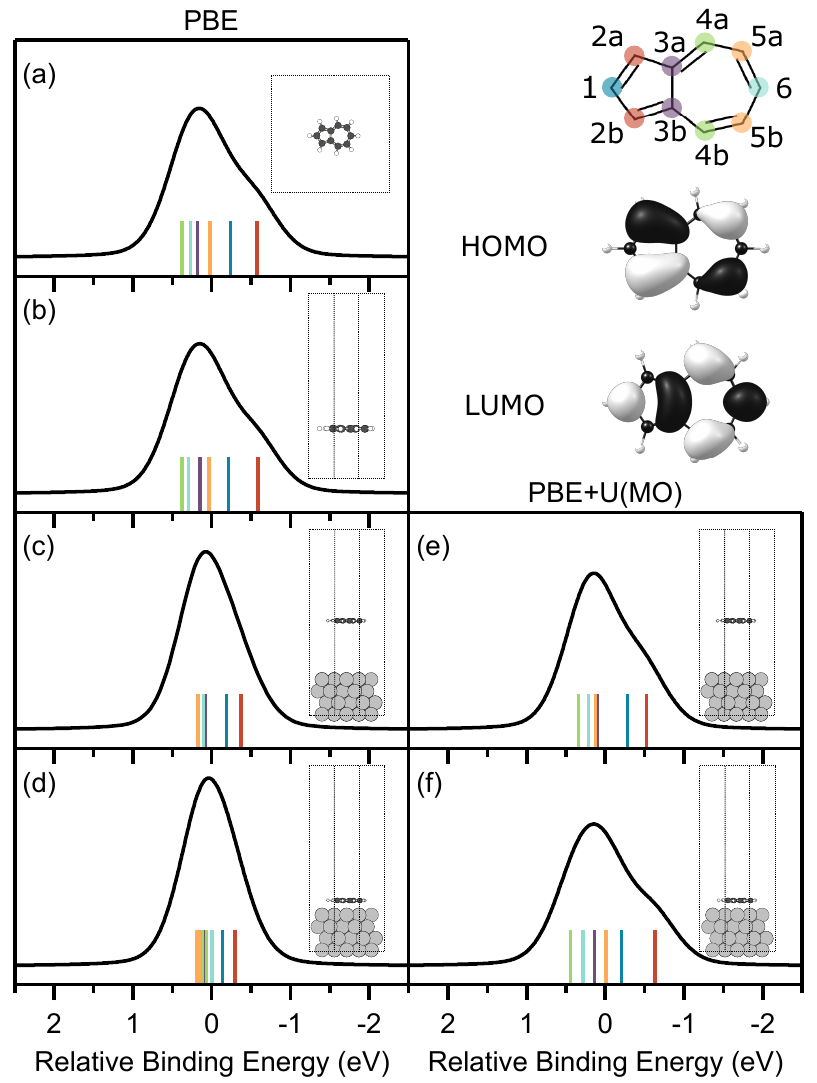}
    \caption{DFT simulated XP spectra of azulene in different configurations with individual atom shifts shown as coloured lines. All spectra have been shifted to align their respective centres of gravity. The colours correspond to the highlighted atoms in the structure shown in the top right corner along with the visualisations of the highest occupied molecular orbital (HOMO) and lowest unoccupied molecular orbital (LUMO). (a) Azulene in the gas-phase, (b) periodic free standing overlayer of Az, (c) Az/Ag(111) with the molecule 12~\AA{} away from the surface, (d) equilibrium geometry of Az/Ag(111). The spectra in (e) and (f) used the same structures as (c) and (d) but include the PBE+U(MO) correction.}
    \label{fig:2}
\end{figure}

Fig.~\ref{fig:2}a shows the XP spectrum of gas-phase azulene and \ref{fig:2}d the spectrum for the equilibrium structure of azulene adsorbed on Ag(111), both including the contributions of all atomic species. All shifts in the adsorbed species are much more closely spaced than in the gas-phase and more closely resemble the case of gas-phase naphthalene rather than azulene. The simulated spectra are shifted so that their respective centres of gravity align. It appears that carbon atom 1, which is located at the apex of the 5-membered ring of azulene is least affected by the changes whereas all atoms with higher binding energies (BEs) located at the 7-membered ring are shifted to smaller BEs and lie closer together in the adsorbed case. Simultaneously, carbon atom 2 on the 5-membered ring is shifted to higher BEs. The result is that all BEs lie within a range of less than \SI{1}{\electronvolt} and the resultant XP spectral envelope constitutes a narrow peak. It is clear that the azulene on Ag(111) slab model predicts a strong change in chemical environment on some carbon atoms, which is not present in the experimental data. 

Possible causes for the discrepancy with experiment can lie in model errors related to structure, electrostatics, or electronic structure. In previous works we have shown that the lateral and vertical adsorption geometry of the molecule is in excellent agreement with experimental evidence from LEED, STM, AFM, and XSW measurements.\cite{klein_molecular_2019,klein_molecule-metal_2019} We can, therefore, exclude any errors in our simulations that stem from misrepresentation of the adsorption structure. It is also known that the chemical shifts are not very sensitive to minor structural changes, which we have confirmed with test calculations for this system (not shown). 

DFT core-hole calculations based on $\Delta$SCF of periodic structures are known to be prone to electrostatic artefacts that arise from finite size effects.\cite{taucher_final-state_2020} Core-holes are introduced in all periodic repeats of the unit cell and if unit cells are not sufficiently large, core-holes can electrostatically interact with each other, which leads to shifts in the binding energies. These artefacts typically converge relatively quickly with unit cell size and should not be significant in this case.\cite{klein_nuts_2021} Unfortunately, the introduction of a homogeneous background charge in periodic core-hole calculations, leads to electrostatic interactions that introduce absolute BE shifts that converge slowly with unit cell size.\cite{taucher_final-state_2020,zojer_impact_2019} However, we have recently shown that this typically does not affect relative BE shifts between different species once reasonable unit cell dimensions are established.\cite{klein_nuts_2021} Kahk et al. have recently proposed an approach that remedies these electrostatic effects and enables the prediction of absolute BEs for periodic structures.\cite{kahk_core_2021}

Fig.~\ref{fig:2}b shows the simulated spectrum of the free standing overlayer of azulene in the periodicity of the surface slab model. The resulting XP spectrum is virtually unchanged compared to the gas-phase and correctly exhibits a shoulder at low BE. The fact that electrostatics are not the cause of the discrepancy with experiment is further corroborated by our XPS calculations on cluster cut-outs of the azulene on Ag(111) structure presented in Fig. S1 (ESI\dag). Here we see that, even without the use of periodic boundary conditions, the C 1s XPS of adsorbed azulene lacks a shoulder at low energy.

Having excluded the effects of structure and electrostatics, we turn our attention to the electronic structure. It is reasonable to assume that the misrepresentation of the XP spectrum could arise from a misrepresentation of charge transfer between molecule and surface. In the top right corner of Fig.~\ref{fig:2}, the HOMO and LUMO frontier orbitals of azulene are depicted. We can see that the HOMO has more amplitude located at the 5-membered ring, while the LUMO has more probability amplitude located on the 7-membered ring. Because the HOMO is occupied and the LUMO is not, the net charge on the carbon atoms in the 5-membered ring is negative whereas the 7-membered ring is partially positively charged, giving rise to the substantial dipole moment of the molecule (see Table S1 for net atomic charge and dipole calculations, ESI\dag). Due to this inhomogeneous charge distribution pattern, charge transfer into the LUMO will lead to more charge being located at the 7-ring and a more homogeneous charge distribution in the anion with a smaller total dipole moment (seen in the calculated dipole moments in Table S1, ESI\dag), which will affect the BE differences of the carbon atoms. We have performed gas-phase calculations of the XPS spectra of neutral and anionic azulene, which are presented in Fig. S2 (ESI\dag). As can be seen, the consequence of adding an electron to azulene is that the carbon BEs become more closely spaced, leading to a narrow XPS signature and to the removal of the shoulder in the spectrum. We suspect that  the shift of the carbon atoms 1,3,4 and 6, where the LUMO is localised, is most responsible for the change in peak shape. 
The occupation of the LUMO is likely also the cause for the removal of the XPS shoulder in the case of azulene adsorbed on Cu(111) where significant surface-molecule charge transfer could be proved.\cite{klein_molecular_2019,klein_molecule-metal_2019}

\begin{figure}[h]
    \centering
    \includegraphics{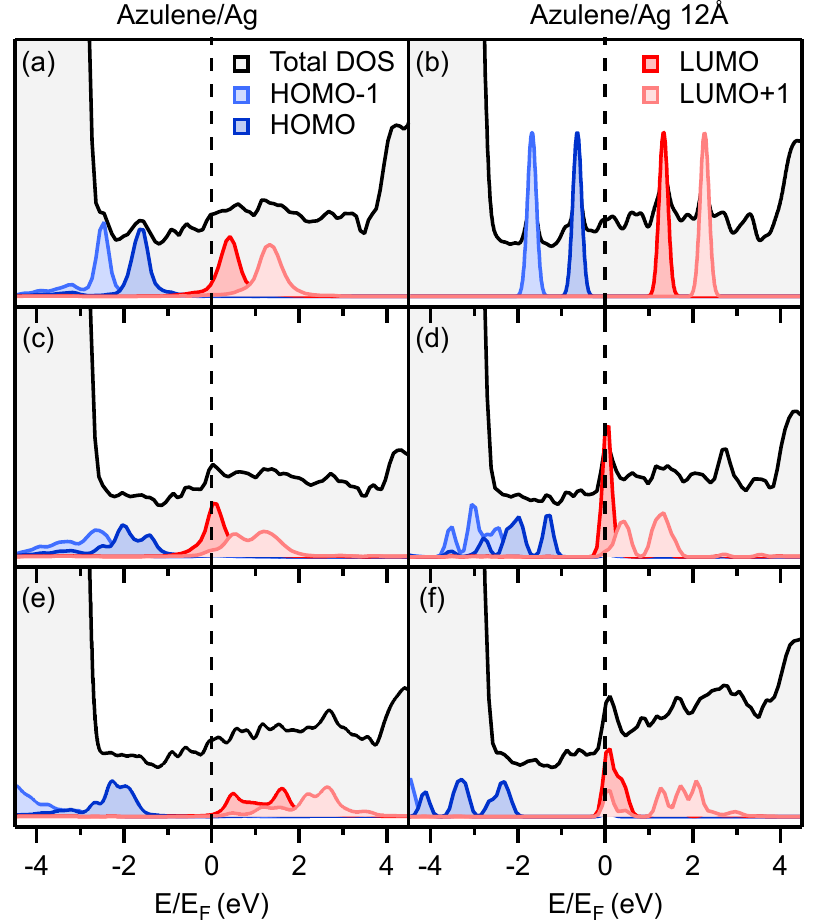}
    \caption{Total density of states and MODOS of structures of azulene adsorbed on a Ag(111) surface (left) and with the azulene molecule 12~\AA{} away from the metal surface (right). (a) and (b) show the results of a ground-state system. (c) and (d) are the core-hole-excited state MODOS. For this plot, the MODOS for all possible core-hole excited carbon species in the molecule have been calculated and summed up. (e) and (f) are the summed up excited core-hole MODOS with the +U(MO) correction. Total DOS is shown by the black line with gray shading. Occupied orbitals are coloured in blue and unoccupied orbitals in red. MO contributions have been scaled for for ease of viewing.}
    \label{fig:3}
\end{figure}

However, we know from our previous work that the charge transfer between molecule and surface for azulene on Ag(111) as predicted by DFT is very small and only amounts to \SI{0.01}{e} as predicted by Bader charge analysis.\cite{klein_molecule-metal_2019} Evidence for this is provided by the MO projected DOS for the adsorbed molecule presented in Fig.~\ref{fig:3}a. All occupied gas-phase MOs are located clearly below the Fermi level; LUMO and LUMO+1 are fully located above the Fermi level. Yet, the simulated XP spectrum is consistent with what we find for a charged gas-phase azulene molecule.

The MODOS in Fig.~\ref{fig:3}a also shows that the molecule is only weakly hybridised with the surface as all MO features are energetically well localised. However, if the issue is due to overly strong chemical interaction and hybridisation rather than charge transfer, then a Gedankenexperiment where we lift the molecule from the metal surface should be able to fully restore the XPS spectrum as predicted by the free standing overlayer shown in Fig.~\ref{fig:2}b. The result for a calculation where the molecule is \SI{12}{\angstrom} away from the surface is shown in Fig.~\ref{fig:2}c and it is puzzling: The XP shoulder at low BE is not restored. While there are some differential shifts within the peaks, they are not of sufficient magnitude to restore the expected peak shape observed in experiment or in the simulated gas-phase or free standing overlayer spectra.

We have established that the simulation error is related to the presence of the surface and consistent with charge transfer into the molecule, yet no evidence for such transfer exists in the ground-state DFT calculation for the adsorbed or lifted molecule (see Fig.~\ref{fig:3}a and \ref{fig:3}b). We therefore, turn our attention to the core-hole calculations that we use to simulate the BE shifts. To simulate a spectrum for $N$ number of carbon atoms, we perform $N$ independent DFT calculations where a 1s core-hole is introduced for each carbon atom in the molecule. The presence of the core-hole leads to orbital relaxation effects and screening of the effective potential that are captured in the total energy that we use to calculate the BE shift. Core-hole relaxation effects can lead to the lifting of orbital degeneracies and orbital realignment that contribute to screening or descreening of the relevant carbon atom. As shown in Fig.~\ref{fig:3}c and \ref{fig:3}d where we visualise the calculated MODOS for azulene on Ag(111) in the presence of the core-hole, the core-hole relaxation effects on the MODOS of this system are substantial. Most importantly, the LUMO is shifted to lower energies to the point that it becomes Fermi level pinned. This is the case for the adsorbed molecule and the molecule at \SI{12}{\angstrom} distance, where charge transfer is physically impossible due to the lack of orbital overlap.

In summary, the root cause of the discrepancy between experiment and simulation is therefore an artificial charge transfer between surface and molecule that arises during the core-hole calculations. The changes in effective potential as described by the underlying approximate xc functional lead to Fermi level pinning even at a large distance from the surface. This is in contradiction to experimentally measured XP spectra, which are not consistent with a significant charge transfer into azulene upon adsorption. In the following, we will apply a correction to our calculations that will demonstrate that this is actually an artefact of the self-interaction error in the xc description within our DFT calculations. 

\subsection{Molecular orbital based self interaction correction: DFT+U(MO)}

As can be seen in Fig.~\ref{fig:3}c and \ref{fig:3}d, by removing a core-electron from the C 1s orbital, the effective potential on the carbon atom becomes more attractive, which reduces the energy of electronic states with probability on this carbon atom. This includes most electronic states. Therefore, the fact that the LUMO orbital of azulene becomes Fermi-level-pinned in the core-hole excited state, is likely related to the error in the electron affinity and ionisation potential, \textit{i.e.} the underestimation of the HOMO-LUMO gap. This underestimation arises from the many-electron self-interaction error of semi-local xc functionals, such as the here employed PBE functional. To explore this point, we artificially adjust the gap between the HOMO and LUMO state of the azulene molecule with a penalty functional that we add to PBE. We call this approach PBE+U(MO) as it is similar to the use of atomic-orbital-based self-interaction corrections in +U methods,\cite{anisimov_band_1991,dudarev_electron-energy-loss_1998,pickett_reformulation_1998,petukhov_correlated_2003} but the projector states target gas-phase MOs. This approach has previously been implemented in the CASTEP code.\cite{maurer_first-principles_2014,muller_interfacial_2016}

The DFT+U(MO) energy functional\cite{maurer_first-principles_2014} is augmented with an additional penalty term that leads to an orbital-dependent potential that increases or reduces the energy of the MO under consideration:

\begin{equation}\label{eq:plus_U}
    \hat{V}_{c} = \frac{U_c}{2} \ket{\phi_c}\bra{\phi_c}
\end{equation}

In equation~\ref{eq:plus_U}, $\phi_c$ refers to the wave function of the reference MO that is to be constrained. The potential is applied in each step of the self-consistent-field algorithm and leads to the variational optimisation of the Kohn-Sham orbitals under this constraint. The choice of U values for each constrained MO, $c$, is \textit{ad hoc} and needs to be manually provided. In our calculation, we apply a constraint to the HOMO, the LUMO, and LUMO+1 orbitals. We identify the values of U for the PBE+U(MO) calculation by searching the space of possible U values that best match the HOMO-LUMO gap predicted by the PBE0 functional\cite{adamo_toward_1999} for the gas-phase azulene molecule in a supercell box. As shown in Table~\ref{tab:1}, we can match the HOMO-LUMO gap of the PBE0 functional with the PBE approach with a choice of $U_{\mathrm{HOMO}}=+2.00$~\si{\electronvolt}, $U_{\mathrm{LUMO}}=+5.25$~\si{\electronvolt}, and $U_{\mathrm{LUMO+1}}=+5.30$~\si{\electronvolt}. Several combinations of parameters have been explored including choices with and without application of a constraint to the LUMO+1 orbital. In cases where we do not constrain the LUMO+1, the orbital ordering is affected and the LUMO+1 becomes Fermi level pinned. The same was applied for the gasphase naphthalene molecule for the data presented in the ESI\dag{} where the values were $U_{\mathrm{HOMO}}=+2.00$~\si{\electronvolt}, $U_{\mathrm{LUMO}}=+5.50$~\si{\electronvolt}, and $U_{\mathrm{LUMO+1}}=+5.55$~\si{\electronvolt}.

\begin{table}[]
    \caption{DFT calculated gap between the HOMO and LUMO of an azulene molecule in a 20~\AA{} periodic vacuum cube performed with either the PBE0 or PBE exchange correlation functional. PBE+U(MO) value is from a calculation where the positions of the HOMO and LUMO and LUMO+1 were shifted by +2.00, +5.25 and +5.30 eV, respectively}
     \label{tab:1}
    \begin{tabular*}{0.48\textwidth}{@{\extracolsep{\fill}}lc} 
    \hline
    xc Functional & HOMO-LUMO Gap (eV)  \\
    \hline
    PBE0      & 3.60 \\
    HSE06     & 2.84 \\
    PBE       & 2.07 \\
    PBE+U(MO) & 3.60 \\
    \hline
    \end{tabular*}
\end{table}

We test the validity of the PBE+U(MO) approach by first applying it to the molecule at \SI{12}{\angstrom} distance from the metal surface shown in Fig.~\ref{fig:2}e. The method indeed predicts the correct shape of the XPS in agreement with experiment. The individual carbon BEs and their relative shifts are in fair agreement with what is found for the case of the free standing overlayer shown in Fig.~\ref{fig:2}b. Some of the shifts deviate, particularly the relative BE shift between carbon 1 and 2 in the 5-membered ring. This may be related to the fact that the +U(MO) correction does not fully remove the artificial charge transfer into the molecule for this geometry, as can be seen in the MODOS for the core-hole excited state calculated with PBE+U(MO) shown in Fig.~\ref{fig:3}f. This goes to show that even orbital energies equivalent to the hybrid functional level with PBE0 may not be fully sufficient to remove all artefacts.

In Fig.~\ref{fig:2}f, we show the PBE+U(MO) XP spectrum for the metal-adsorbed geometry, which is in good agreement with the experimentally measured spectra for the monolayer and the multilayer of azulene adsorbed at Ag(111). Particularly noteworthy is the fact that the individual ordering of the carbon atoms in terms of their BE is fully preserved upon adsorption of the molecule. This is not unexpected as the molecule is only weakly hybridised and adsorbs in a physisorbed state at about \SI{3.1}{\angstrom} from the metal surface.\cite{klein_molecule-metal_2019}

This is the same scenario as for naphthalene on Ag(111), where the multilayer, monolayer, and gas-phase XP spectra of the molecule are virtually identical. We note that the spectral prediction for naphthalene on Ag(111) may also suffer from errors associated with the xc approximation, but similarity between the BEs of all the carbon atoms makes it impossible to use the experiment as meaningful benchmark reference. In fact, naphthalene on Ag(111) suffers from the same artificial charge transfer issue as azulene as can be seen in Fig. S3b (ESI\dag). But this issue does not measurably materialise, because due to the homogeneous electronic structure of naphthalene, the charge transfer shifts the contributions of all atoms equally and does not introduce a relative shift as can be seen on in Figure S5 (ESI\dag). In Fig. S3c and S4 (ESI\dag) we compare the results of the PBE+U(MO) approach for the naphthalene on Ag(111) XP spectrum with the regular PBE result and find that the total convoluted spectra are virtually indistinguishable. In contrast, the artificial charge transfer into azulene predominantly affects the shifts of the carbons where the LUMO has significant probability amplitude (see Figure S5, ESI\dag), which changes the shape of the spectrum.

We further stress that this approach is not a universal solution to charge transfer artefacts that can arise from self-interaction error, but rather an \textit{ad hoc} correction to exemplify the problem. The penalty potential applied to MO gas-phase reference orbitals is also only reasonable in cases where the MO gas-phase orbitals remain meaningful representations of the electronic structure, so where the molecule is only weakly hybridised with the surface.

Lasting solutions to this artefact require either the use of better exchange correlation functionals with non-local exchange and correlation description or the use of excited-state methodology beyond simple $\Delta$SCF-based core-hole constraints. We note that even the use of range-separated hybrids such as HSE06\cite{heyd_hybrid_2003} does not remedy the shoulder problem for the cluster cut-out we studied in Fig. S1 (ESI\dag), however, the HOMO-LUMO gap predicted by HSE06 is also considerably lower than that of PBE0. Unfortunately, bare hybrid xc functionals such as PBE0 cannot simply be applied to the study of metals as they do not provide a reliable prediction of the metal electronic structure, but optimally tuned range-separated hybrid functionals with custom optimised parameters for individual systems may be able to address the problem. \cite{egger_reliable_2015,wruss_distinguishing_2018}

\subsection{Implications of charge transfer artefacts for NEXAFS predictions}

\begin{figure}[h]
    \centering
    \includegraphics{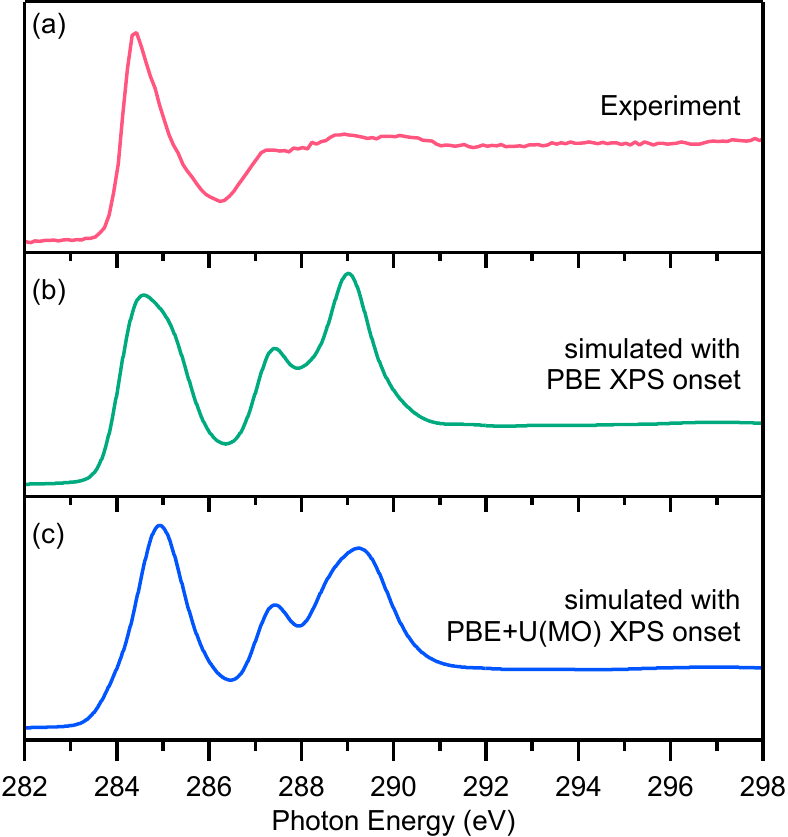}
    \caption{Comparison of the experimentally measured and simulated  NEXAFS data for azulene/Ag(111). (a) experimental NEXAFS spectrum. (b) and (c) show NEXAFS simulations using the $\Delta$IP-TP method with two different onset corrections to sum up the spectra. (b) uses the XPS binding energies obtained from a $\Delta$SCF calculation with PBE and (c) from a $\Delta$SCF calculation with PBE+U(MO). The simulated spectra have been shifted to align with the experimental energy scale. For both experimental and simulated data, a 25$^\circ$ incidence angle was chosen. Experimental data taken from Ref.\cite{klein_molecule-metal_2019}.}
    \label{fig:4}
\end{figure}

The role of xc errors and the resultant charge transfer artefacts in XPS simulation of metal-organic interfaces is a cautionary tale as this can lead to the misinterpretation of spectra and wrong conclusions on the surface chemistry. However, this error can also further affect the simulation of K-edge NEXAFS spectra at metal-organic interfaces, which are highly valuable for the interpretation of the surface-adsorption-induced changes in electronic structure and also the molecular orientation at the surface.\cite{stohr_nexafs_1992,breuer_characterization_2015}

While the most common approach to XPS simulations at the DFT level is the use of $\Delta$SCF in combination with the maximum overlap method,\cite{besley_self-consistent-field_2009} x-ray absorption is commonly simulated with the transition potential (TP) method\cite{triguero_calculations_1998} based on Slater's transition state approach.\cite{slater_statistical_1972} We have recently discussed this approach and its technical and numerical aspects in detail.\cite{klein_nuts_2021} In its realisation in periodic plane wave codes, the core-hole relaxation effects on the valence electronic states are simulated via the introduction of a half core-hole and a single transition potential calculation of the Kohn-Sham eigenvalues. As the atomic nucleus is represented by a pseudopotential in this approach, the ionisation potential associated with the removal of an electron from the 1s state is then corrected by shifting the spectrum with respect to the BE of the 1s state of the relevant carbon atom calculated with the $\Delta$SCF method, we call this approach $\Delta$IP-TP.\cite{klein_nuts_2021,mizoguchi_first-principles_2009} Therefore, prediction errors in the XPS propagate into errors in the NEXAFS predictions as changes in relative BE shifts will lead to the misalignment of excitations that arise from different core-states into the same valence states.

In Fig.~\ref{fig:4} we show the experimentally measured K-edge NEXAFS spectrum of azulene adsorbed on Ag(111)\cite{klein_molecule-metal_2019} against the $\Delta$IP-TP prediction corrected by PBE XPS onsets (in Fig.~\ref{fig:4}b) and PBE+U(MO) onsets (in Fig.~\ref{fig:4}c). We can see that the first resonance of the NEXAFS spectrum based on the U-corrected XPS energies becomes narrower and the broad shoulder present at about \SI{285}{\electronvolt} in the PBE-only prediction is reduced. Note that we use the conventional PBE functional for the TP calculations of the valence state relaxation and not the +U(MO) approach, because the orbital ordering and spacing of the unoccupied states will likely be affected by the +U(MO) correction. This is due to the fact that we have only selected U values under the criterion to reproduce the PBE0 HOMO-LUMO gap, but not to correctly capture the optical 1s$\rightarrow$LUMO+X excitations of the system.

\section{Conclusion}
When calculating core-level spectra for metal-organic hybrid interfaces it is important to take into consideration various errors that can affect both the absolute binding energies and relative shifts. Binding energies are difficult to predict on an absolute energy scale with $\Delta$SCF and conventional exchange-correlation approximations in DFT. However, relative shifts between individual atoms are often less influenced by intrinsic DFT errors. In most cases, when studying organic molecules adsorbed at surfaces, the overall shape of the spectrum can therefore be predicted accurately. This does not hold true for azulene on Ag(111), where spurious charge transfer between the metal and the molecule is introduced by the presence of the core-hole. As a consequence, the relative shifts between the C 1s binding energies of the different carbon atoms are changed and the predicted shape of the XP spectrum is too narrow and lacks a shoulder that is present in the experiment. The problem is seen across different DFT software packages and occurs for GGA, meta-GGA, and range separated hybrid functionals, both in periodic and aperiodic simulations. 

While this artificial charge transfer is likely present in many molecule-metal interface calculations and we also find it to be present in our simulations for naphthalene on Ag(111), only azulene on Ag(111) seems to be particularly sensitive to this effect. The shoulder in the C 1s XPS of azulene originally arises from the non-alternant topology of the molecule which leads to an inhomogenous charge distribution, which result in unusually large relative binding energy shifts within a $\pi$-conjugated organic molecule. The spurious charge transfer into the molecule that is caused by the core-hole screening and the underestimation of the HOMO-LUMO gap leads to a more homogeneous charge distribution in the molecule, which also leads to smaller shifts in binding energies of different carbon atoms. The spurious charge transfer is ultimately caused by DFT self-interaction error and we show that it can be addressed by applying a penalty functional method, DFT+U(MO), to either increase or decrease the energy of unoccupied and occupied molecular orbital states. By applying this \textit{ad hoc} correction we are able to predict the correct shape of the XP spectrum of azulene on Ag(111) and furthermore also improve the prediction of the NEXAFS spectrum.

Artificial charge transfer caused by core-hole screening is likely to be common in the simulated core-level spectra of metal-organic interfaces, but more often than not it will be invisible as the experimental resolution of individual carbon shifts will be insufficient. Systems similar to azulene with inhomogeneous charge distributions may be more strongly affected by this problem and caution is advised when performing and interpreting core-level simulations in these cases. Potentially problematic systems include molecules with intramolecular charge-transfer and low-lying conduction states, such as donor-acceptor polymers or polythiophenes which are commonly used in organic electronics devices. While Many Body Perturbation Theory remains out of reach for large-scale metal-organic interface models, improved exchange-correlation functionals will be required to address the cause of this problem. Due to the difficulty of addressing metallic systems with hybrid functionals, currently only optimally tuned range separated hybrid functionals may be able to address this problem.\cite{egger_reliable_2015,wruss_distinguishing_2018}

\section*{Acknowledgements}
This work was funded by the UKRI Future Leaders Fellowship programme (MR/S016023/1), the Walter-Benjamin programme of the Deutsche Forschungsgemeinschaft (DFG, KL 3430/1-1), and the EPSRC-funded Centre for Doctoral Training in Molecular Analytical Sciences (MAS CDT, EP/L015307/1). We acknowledge computational resources provided by the Scientific Computing Research Technology Platform of the University of Warwick, the EPSRC-funded HPC Midlands+ computing centre (EP/P020232/1) and on ARCHER2 (https://www.archer2.ac.uk/) via the Materials Chemistry Consortium (EP/R029431/1). 

\section*{Competing Interests}
The authors declare no competing interests.


\section*{Data Availability}
The input and output files of the electronic structure calculations reported in this work have been deposited to the NOMAD repository and can be accessed here: DOI will be included upon acceptance by publisher.


\bibliography{reference} 
\bibliographystyle{rsc.bst} 

\end{document}


\maketitle

\noindent$^a$ Department of Chemistry, University of Warwick, Gibbet Hill Rd, Coventry, CV4 7AL, United Kingdom \\

\noindent$^b$ MAS CDT, Senate House, University of Warwick, Gibbet Hill Rd, Coventry, CV4 7AL, United Kingdom \\

\noindent$^c$ Diamond Light Source, Harwell Science and Innovation Campus, Didcot, OX11 0DE, United Kingdom

\pagebreak

\begin{figure}[h]
    \centering
    \includegraphics{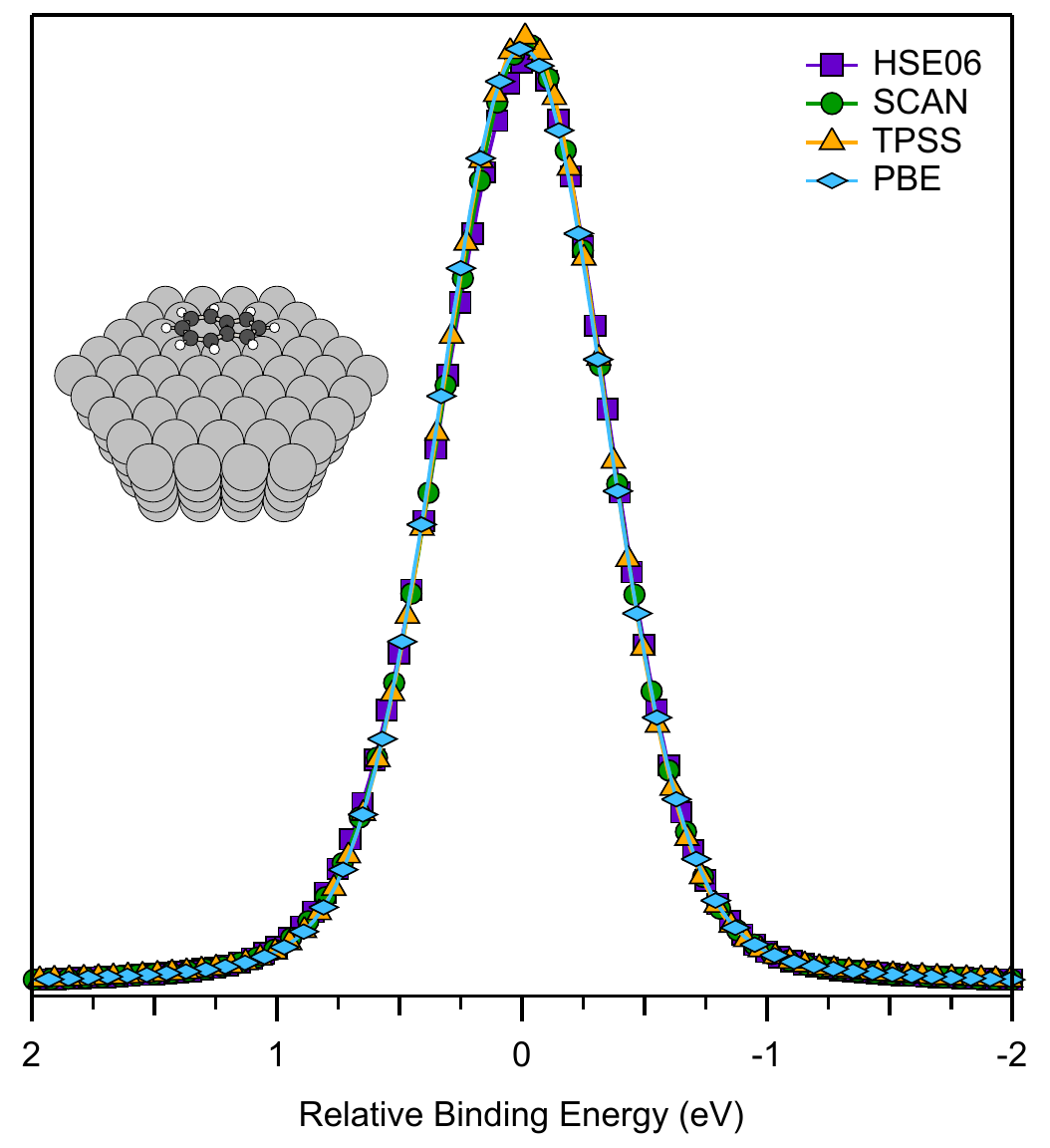}
    \caption{Simulated XP spectra, calculated using the $\Delta$SCF method, of azulene on Ag(111) using an aperiodic cluster structure shown alongside. Calculations were performed using the all-electron DFT software package FHI-aims for three different exchange correlation functionals, HSE06 shown in purple, SCAN in green and TPSS in yellow. HSE06 was calculated using the \textit{force\_occupation\_basis} keyword to constrain the core-hole whilst SCAN and TPSS utilised the \textit{force\_occupation\_projector} keyword. All spectra are aligned to the average shift of the individual atoms.}
    \label{fig:SI_XPS}
\end{figure}

\begin{table}[h]
    \centering
    \caption{Mulliken and Hirshfeld charges and dipole moments calculated for gas-phase azulene using FHI-aims. Shown are the joint partial charges of the carbon and hydrogen atoms for each symmetry inequivalent CH group. All charges are in units of e, the dipole moments are in Debye. }
    \begin{tabular}{cc|SSSSSSS|c}
    \multicolumn{2}{c|}{Carbon} & 1 & 2 & 3 & 4 & 5 & 6 & Sum & Dipole Moment \\
    \hline
    \multirow{2}{*}{Mulliken} & Neutral & 0.00 & -0.12 & 0.10 & 0.01 & -0.02 & 0.02 & 0.00 & 1.03  \\
         & Anion & -0.15 & -0.31 & 0.07 & -0.28 & -0.18 & -0.14 & -0.01 & 0.82 \\
    \hline
    \multirow{2}{*}{Hirshfeld} & Neutral & 0.00 & -0.07 & -0.00 & 0.06 & -0.02 & -0.02 & -1.00 & 0.85 \\
         & Anion & -0.14 & -0.23 & -0.10 & -0.18 & -0.18 & -0.13 & -0.96 & 0.67
    \end{tabular}
    \label{tab:SI_charges}
\end{table}

\pagebreak

\begin{figure}
    \centering
    \includegraphics{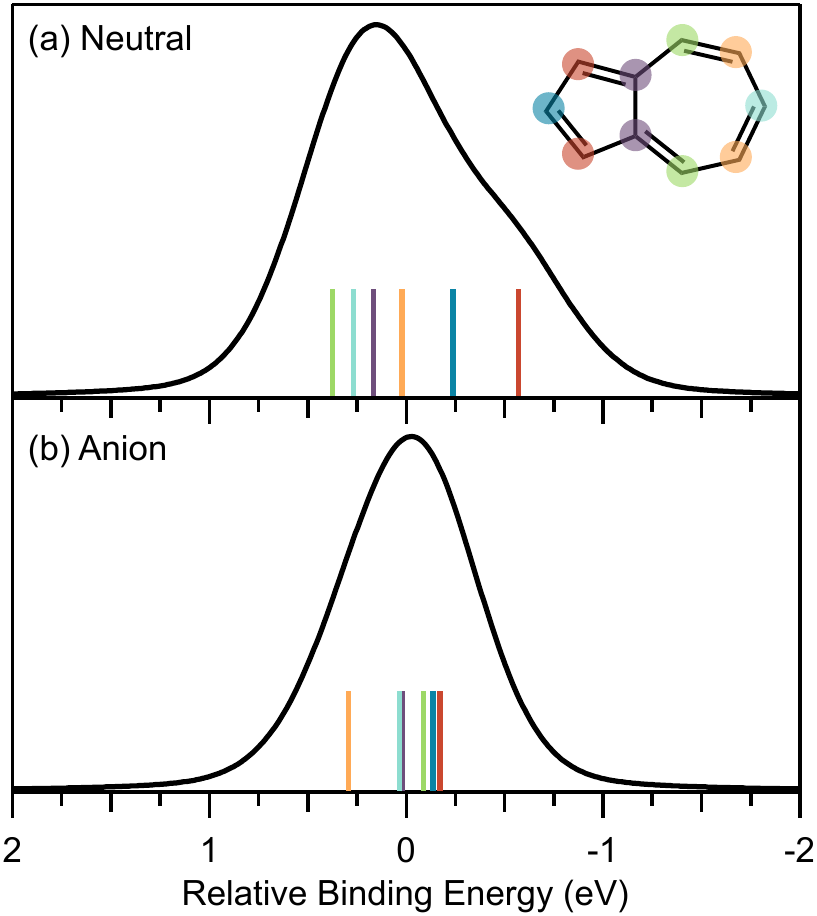}
    \caption{DFT simulated XPS of azulene in a gasphase aperiodic structure, performed with the all-electron software package FHI-aims with the indivdual atom shifts shown as coloured lines according to the labelled structure. Spectra are calculated using the $\Delta$SCF method with (a) keeping the molecule overall charge neutral or (b) with the molecule left in an anionic state. Spectra have been aligned to the average shift of the individual atom shifts.}
    \label{fig:SI_AzGas}
\end{figure}

\begin{figure}[h]
    \centering
    \includegraphics{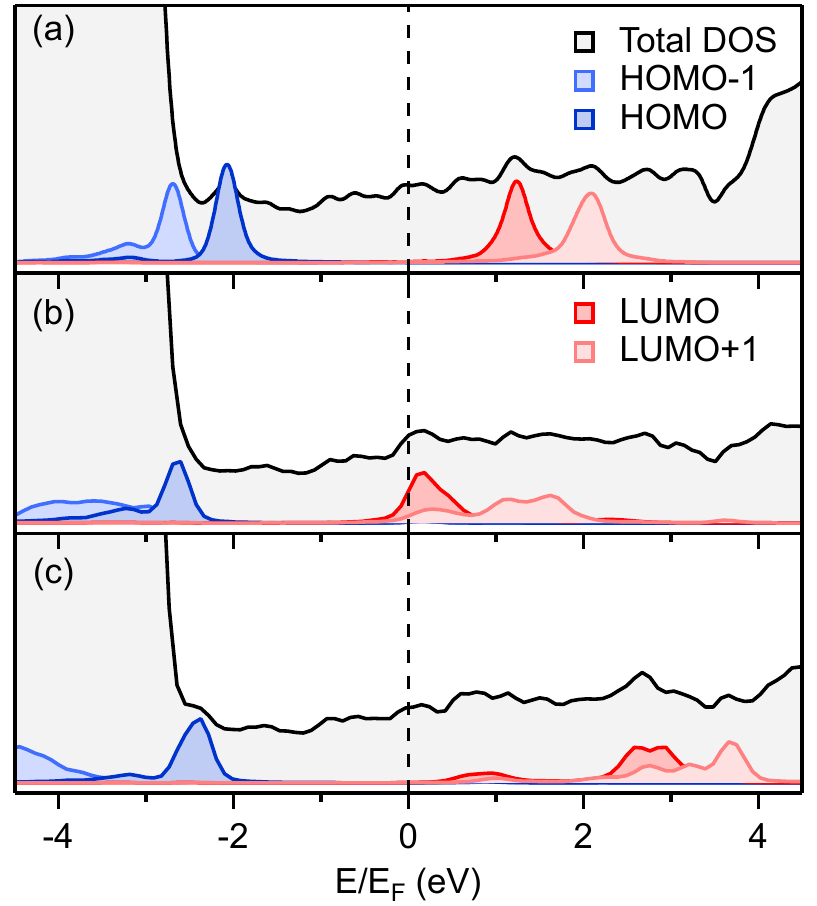}
    \caption{Total density of states and MODOS of naphthalene on Ag(111). (a) shows the DOS in the ground state, (b) in the core-hole excited state with all individual orbitals summed up and (c) the core-hole excited state with the DFT+U(MO) correction applied. Total DOS is shown by the black line with gray shading. Occupied orbitals are coloured in blue and unoccupied orbitals are coloured in red. Orbital contributions have been scaled for ease of viewing.}
    \label{fig:SI_MODOS}
\end{figure}

\begin{figure}[h]
    \centering
    \includegraphics{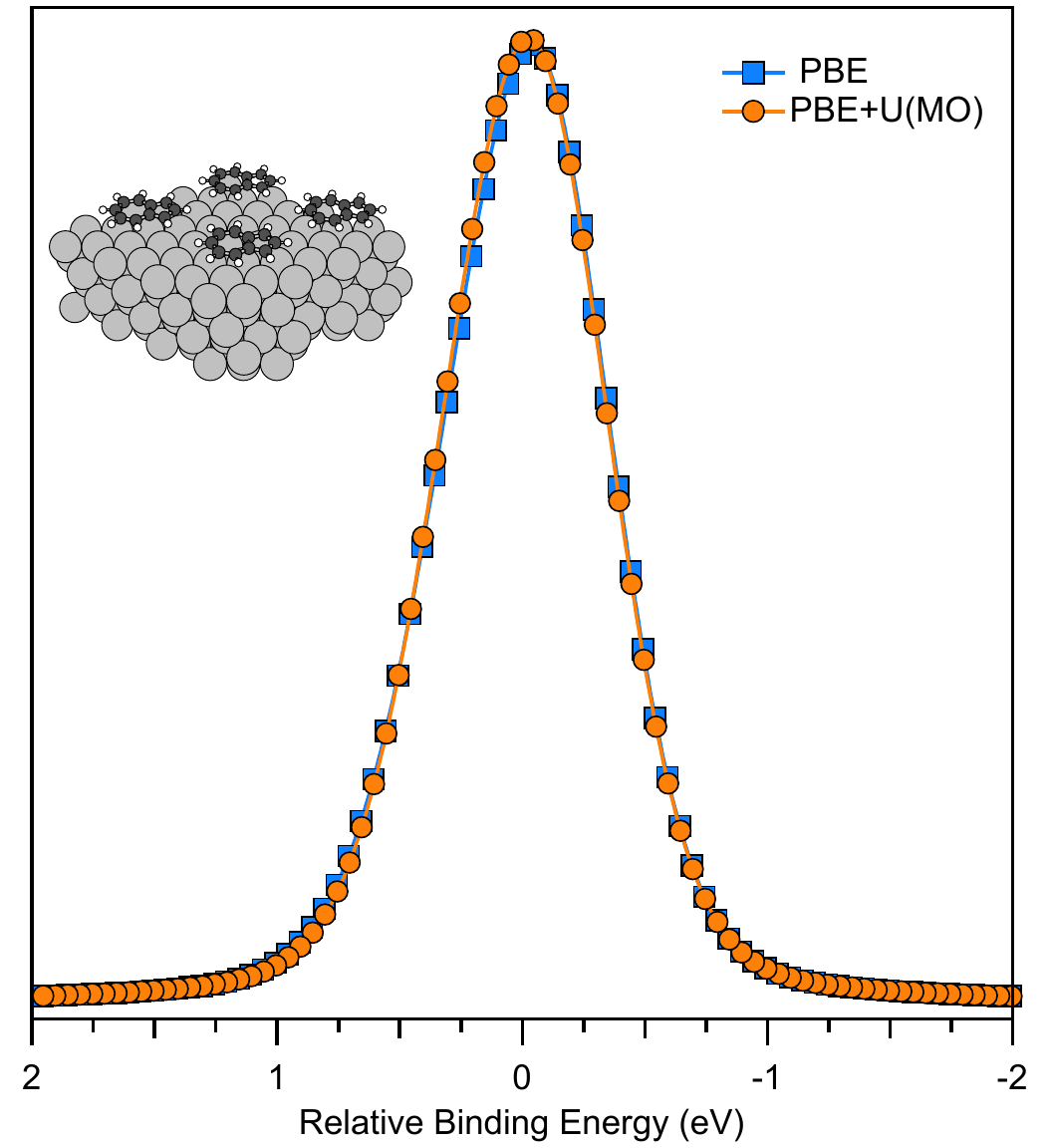}
    \caption{DFT Simulated XP spectra of naphthalene on Ag(111) performed in periodic boundary conditions using the CASTEP code using the $\Delta$SCF method (blue) and with the PBE+U(MO) (orange) to shift specific orbitals. Spectra are aligned to the average shift of the individual atoms.}
    \label{fig:SI_XPS}
\end{figure}

\begin{figure}
    \centering
    \includegraphics{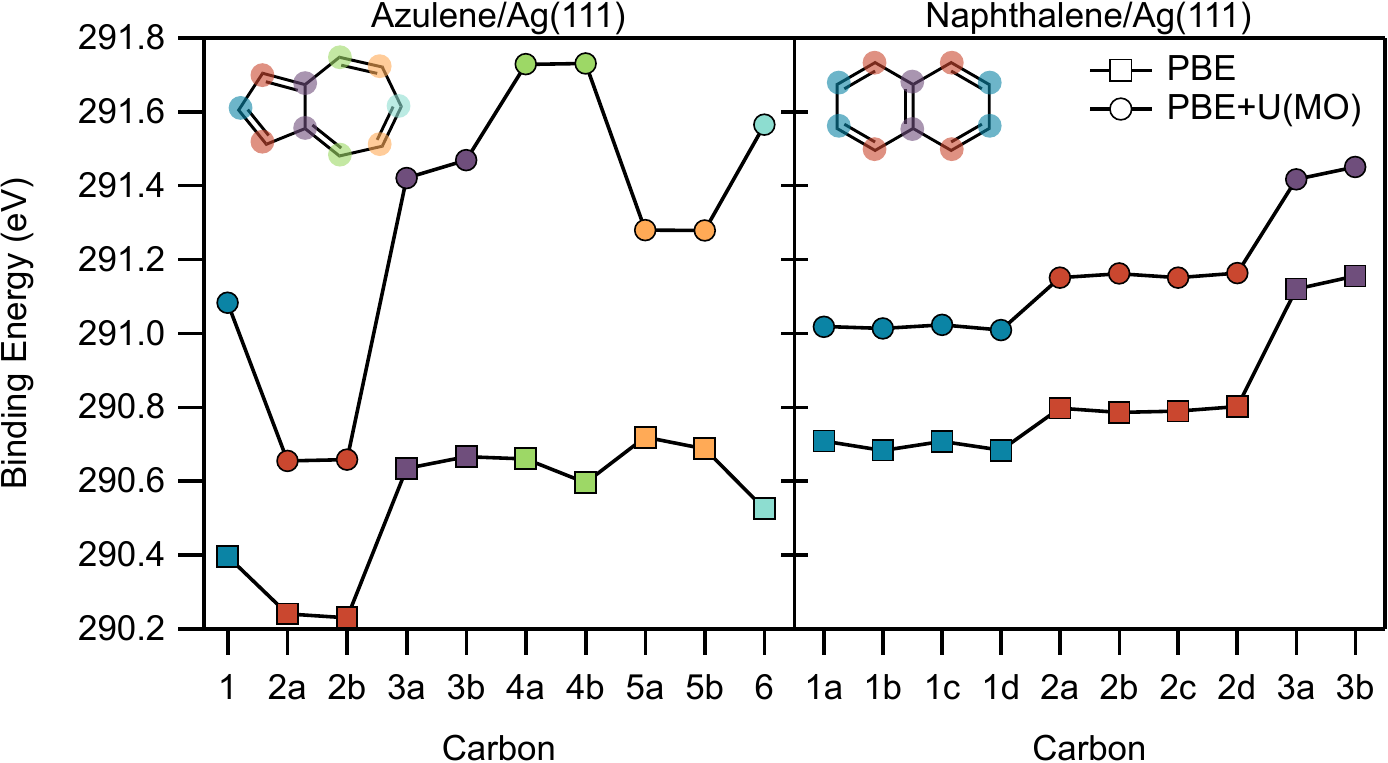}
    \caption{Absolute carbon shifts for each individual carbon atom in azulene (left) and naphthalene (right) adsorbed onto an Ag(111) surface. Squares represent results from a $\Delta$SCF calculation with CASTEP using the PBE functional and circles are results where the PBE+U(MO) correction has been applied.}
    \label{fig:SI_abs_XPS}
\end{figure}

